\documentclass[prl,twocolumn,floatfix,nobibnotes,superscriptaddress]{revtex4}
\usepackage{amsmath}
\usepackage{blindtext}
\usepackage{graphicx}
\usepackage[tight]{subfigure}
\usepackage[english]{babel}
\usepackage{csquotes}
\usepackage{amssymb}
\usepackage{amsmath}
\usepackage{graphicx,textcase}
\usepackage{subfigure}
\usepackage{gensymb}
\usepackage[normalem]{ulem}
\usepackage{color}
\usepackage[citecolor=blue,colorlinks=true]{hyper ref}
\usepackage{xcolor}
\usepackage{hyperref}
\hypersetup{
    colorlinks=true,
    linkcolor=black,
    filecolor=black,      
    urlcolor=black
}

\begin{document}

\title{Spin-polaronics, an emerging technology}
\author{N. Bondarenko}
\affiliation{Division of Materials theory, Department of Physics and Astronomy, Uppsala University, Box 516, 75121 Uppsala, Sweden}
\author{J. Chico}
\affiliation{Division of Materials theory, Department of Physics and Astronomy, Uppsala University, Box 516, 75121 Uppsala, Sweden}
\affiliation{Peter Gr\"unberg Institut and Institute of Advanced Simulation, Forschungszentrum J\"ulich \& JARA, D-52428, J\"ulich, Germany}
\author{Y.O. Kvashnin}
\affiliation{Division of Materials theory, Department of Physics and Astronomy, Uppsala University, Box 516, 75121 Uppsala, Sweden}
\author{A. Bergman}
\affiliation{Division of Materials theory, Department of Physics and Astronomy, Uppsala University, Box 516, 75121 Uppsala, Sweden}
\affiliation{Maison de la Simulation, USR 3441, CEA-CNRS-INRIA-Universit\'{e} Paris-Sud-Universit\'{e} de Versailles, F-91191 Gif-sur-Yvette, France}
\affiliation{L\_Sim, INAC-MEM, CEA, F-38000 Grenoble, France}
\author{N.V. Skorodumova}
\affiliation{Division of Materials theory, Department of Physics and Astronomy, Uppsala University, Box 516, 75121 Uppsala, Sweden}
\affiliation{Multiscale Materials Modelling, Department of Materials Science and Engineering,Royal Institute of Technology, SE-100 44 Stockholm, Sweden}
\author{O. Eriksson}
\affiliation{Division of Materials theory, Department of Physics and Astronomy, Uppsala University, Box 516, 75121 Uppsala, Sweden}
\affiliation{School of Science and Technology, \"Orebro University, SE-701 82 \"Orebro, Sweden}

\begin{abstract}
The static and dynamic properties of spin-polarons in La-doped $CaMnO_3$ are explored theoretically, by means of an effective low energy Hamiltonian. All parameters from the Hamiltoniain are evaluated from first principles theory, without adjustable parameters. 
The Hamiltonian is used to investigate the temperature stability as well as the response to an external applied electric field, for spin-polarons in bulk, surface and as single two-dimensional layers. Technically this involves atomistic spin-dynamics simulations in combination with kinetic Monte Carlo simulations. Where a comparison can be made, our simulations exhibit an excellent agreement with available experimental data and previous theory. Remarkably, we find that excellent control of the mobility of spin-polarons in this material can be achieved, and that the critical parameters deciding this is the temperature and strength of the applied electrical field. We outline different technological implications of spin-polarons, and point to spin-polaronics as an emerging sub-field of nano-technology. In particular, we demonstrate that it is feasible to write and erase information on atomic scale, by use of spin-polarons in $CaMnO_3$. 

\end{abstract}
\pacs{later}

\maketitle
%

\section{I. Introduction}

In the classical formulation, a polaron is a charged carrier localised in a potential well created via self-induced polarisation in the polar crystal \cite{Landau,Pekar}. The polaron concept has been extended to the systems with magnetic interactions, which leads to a new phenomenon, the so-called spin-polaron. Similarly to the classical polaron, the spin-polaron describes a localised charge carrier. However, in this case, the quasiparticle stabilises due to the strong magnetic interaction of an impurity spin and spin states of the host material \cite{Emin1}. In the literature this quasiparticle can be found under a  large variety of terms$-$ spin-polaron \cite{Emin1}, magnetic polaron \cite{Mauger}, ferron \cite{Nagaev1,Nagaev2}, to give some examples. 

\begin{figure}[htb]
\begin{center}
\includegraphics[scale=0.34]{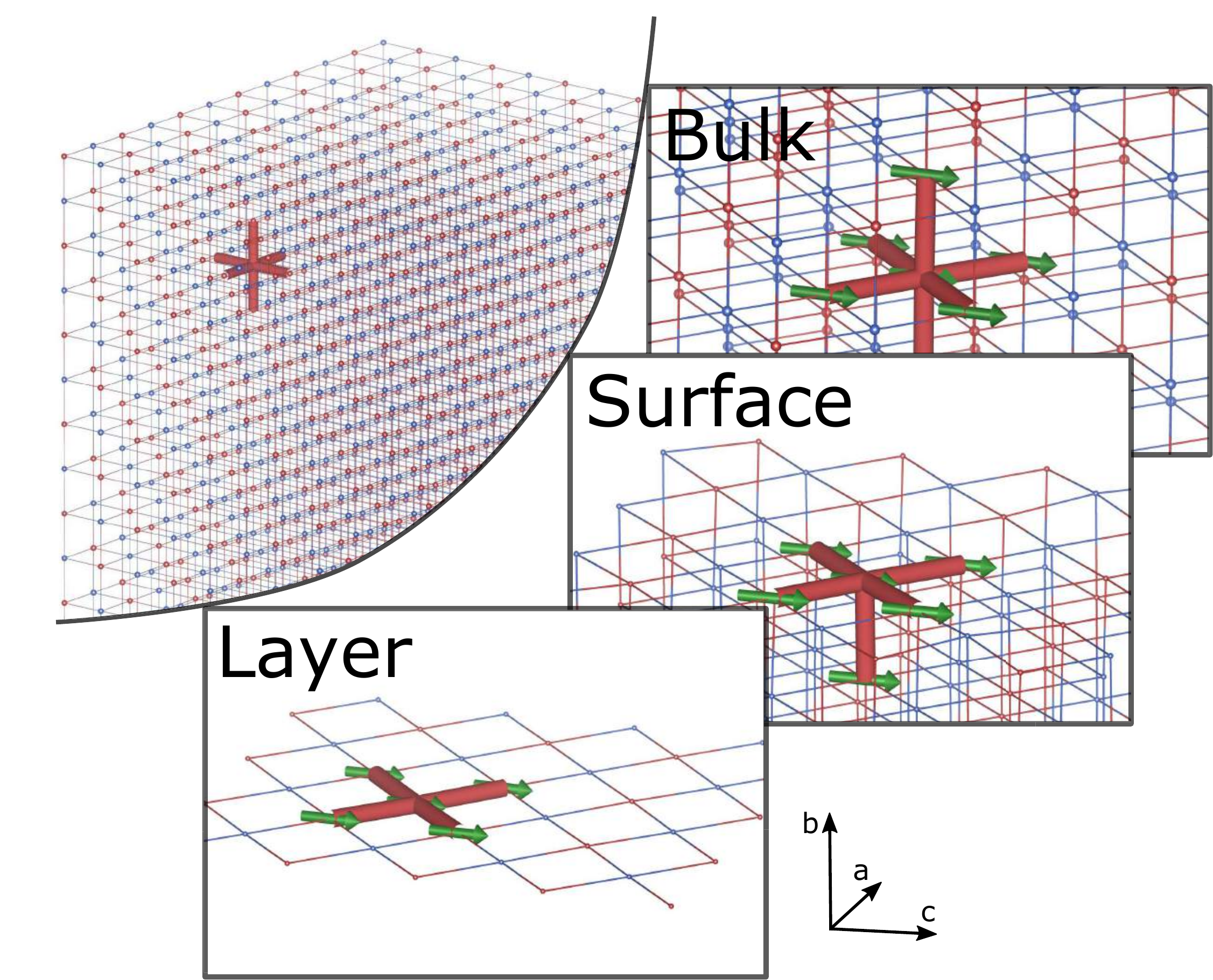}
\end{center}
\caption{\label{structure}(Color online) G-AFM  supercell with one immersed seven-site spin-polaron (marked with red bars, upper left corner). Blue and red sites refer to atoms with magnetisation pointing up and down, respectively. In the insets (right hand part of the plot), we illustrate magnetic clouds (spin-polarons) in a free standing two-dimensional layer, on the surface and in bulk. The direction between nearest neighbour Mn atoms is along $<001>$ and next nearest neighbours is along $<101>$. Also, the polarons are ferromagnetically polarized along the $<101>$ direction, according to Ref.\cite{Ling1}.}
\label{fig1} 
\end{figure}

The seminal work \cite{Gennes} studying the spin-polaron quasiparticle triggered great interest and has been followed by an extensive number of publications employing several theoretical methods \cite{Nagaev2, Emin2, Mauger,Umehara,Kuzemsky}. These studies are not only of academic interest but have a major importance in the understanding of e.g. the magneto-transport and magneto-optical effects \cite{Landwehr}, high-temperature superconductivity  \cite{ Mott1},  colossal magnetoresistance phenomena  \cite{Nagaev3}. 

Magnetic polaron formation is a complex quantum-mechanical phenomenon which encompasses electronic, lattice and magnetic effects and their correlations. A carrier that binds to a self-induced magnetisation, can lower the energy of the system, compared to the unbound configuration. Therefore, a gain in magnetic energy due to a carrier localization outweighs the cost of e.g. temperature fluctuations, kinematic energies of electron hopping and possibly a local lattice deformation. In the antiferromagnetic (AFM) lattice, a spin-polaron stabillisation locally breaks magnetic symmetry suppressing long-range antiferromagnetic interactions, such that instead local ferromagnetic interactions are preferred. In a one-dimensional AFM spin chain, the polaron forms over a FM alignment of N spins \cite{ Gonzalez}. In  2D and 3D systems, magnetic polarons can form with a number of spatial configurations \cite{ Meskine1,Meskine2}.   

Studying spin-polaron dynamics increases the complexity of the physical picture compared to regular polarons. As we discussed above, a spin-polaron in an AFM lattice is associated with a local region that carries a FM spin alignment over a few atomic sites. Therefore a spin-polaron motion will be reflected by how this FM region propagates in the spin subsystem, something we rely on in this investigation. According to the theory of rate processes, the transition rate of this quasiparticle depends on e.g. its size, carrier propagation regime, adiabatic or non-adiabatic behaviour and intrinsic material parameters of the system\cite{Liu,Emin3,Kemeny} .

In this study different aspects of spin-polaron physics are explored in La-doped $CaMnO_3$, an orthorhombic ( Pnma)  semiconductor which stabilises in bipartite  (G-AFM) \cite{ Wollan}  magnetic structure ($T_N$=125K).  The spin-polaron formation in this compound can be described according to the following mechanism. Excess electrons, injected via  La doping, accumulate on Mn($e_g$) orbitals locally increasing the nominal charge of Mn from $4+$ to $3+$. Due to the strong Hund's coupling,  carrier localization on Mn$^{3+}$ ions is possible only via spin-flip events, where the entire atomic moment reverses its nearest neighbour exchange from anti-ferromagnetic to ferromagnetic alignment, such that a local FM region is formed (see Fig.1) \cite{Meskine1,Meskine2,Allen}. In other words, the carrier localisation is suggested to lead to magnetic phase separation governed by an interplay of the  AFM superexchange of the host and a FM double-exchange of FM region \cite{Anderson1,Anderson2, Zener, Goodenough}.

Experimentally, FM-droplets have been determined for La-doped $CaMnO_3$,  in the La concentration range of  $0.01 - 0.10$ \cite{Ling1,Ling2,Wang,Neumeier,Cohn1,Cornelius,Chiorescu,Granado1, Granado2}. Neutron powder diffraction and DC-magnetization techniques reported an average FM-droplet size of about 10  \AA~\cite{Ling2,Granado2}, a magnitude that corresponds to a 7-13 site spin-polaron \cite{Meskine1,Meskine2, Allen}. Interestingly, experimentally observed dynamical properties of magnetic polarons provide agreement with different theoretical models. For instance, Hall-mobility and thermopower analysis reported large polaron in the intermediate coupling regime \cite{Cohn2}. Measurement of the electric conductivity of $CaMnO_3-LaMnO_3$ system \cite{Worledge,Ohtaki,Lan}  are in good agreement with the adiabatic small-polaron model \cite{Mott2,Emin4}.

Theoretically, there were several attempts to study magnetic polarons in $La_xCa_{1-x}MnO{_3}$. In a pioneer ab-initio work it has been shown that the spin-polaron sites exhibit $e_g$ character \cite{Meskine1}. Recently, an ab-initio study proposed a detailed microscopic description of magnetic polarons using DFT+U as well as a hybrid functional for the electronic subsystem\cite{Bondarenko3}. This investigation demonstrated that interactions beyond conventional parametrizations of DFT were needed to obtain $e_g$ localization on the polaronic (Mn) sites. Moreover, the authors of Ref.\cite{Bondarenko3} found that the excess charges mainly localize in a double-exchange active (101) plane. Also, it was shown in this work that the size and intrinsic characteristics of the magnetic polarons strongly depend on the La concentration. At lower concentration ( $x<0.02 $),  spin-polaron formation is driven mainly via magnetic effects. With increasing La concentration, both lattice and magnetic effects start to play a significant role.               

The present study is a multiscale approach that involves ab-initio theory, magnetisation dynamics and kinetic Monte Carlo simulations, to address the dynamical properties of spin-polarons in two and three dimensions (2D and 3D respectively). We start by introducing an effective Heisenberg model for the spin-polaron, corresponding to the $La_x Ca_{1-x}MnO_3$ antiferromagnetic lattice. The material specific exchange interaction parameters incorporated in the Heisenberg model have been extracted from the DFT+U calculations combined with the Liechenstein-Katsnelson-Antropov-Gubanov (LKAG) formalism\cite{Liechtenstein}. To study spin-polaron dynamics, we have evaluated energy barriers of single polaron hoping, using the frame-work of Marcus-Emin-Holstein-Austin-Mott theory (MEHAM), previously introduced for polarons in Refs.\onlinecite{Deskins,Bondarenko1,Bondarenko2}. With these barriers and an effective spin-Hamiltonian that is extended to allow for polaron jumps, via Kinetic Monte Carlo (KMC) simulations, we have evaluated single- and multi-polaron dynamics for a range of temperatures and external electric fields (E-fields). Detailed information about polaron dynamics is presented, both when it comes to its stability with respect to thermal fluctuations, as well as the possibility to utilize these objects in nano-technology, as carriers of information. 

\section {Details of Calculations}

In order to get the insight into the microscopic magnetic properties of the doped CMO, a series of calculations were performed using the "RSPt" code\cite{rspt-book}, based on the full-potential realisation of the linear-muffin-tin-orbital method. 
The calculations were done on a grid of 14$\times$10$\times$14  k-points to ensure the convergence of the exchange integrals. 

The effective exchange integrals ($J_{ij}$) between Mn magnetic moments were extracted by means of the magnetic force theorem\cite{licht-exch,licht-exch2} as implemented in RSPt\cite{Kvashnin}.
We have employed the basis set, containing three types of the basis functions, characterised by different kinetic-energy tails ($\{-0.3, -2.3, -1.5\}$Ry) and all three of them were used to describe each of Mn-$3d$ orbitals. 
In order to perform DFT+$U$ and, consequently, the $J_{ij}$ calculation, these states were projected onto the muffin-tin head of Mn (for details, see e.g. Ref.~\onlinecite{Kvashnin}).
To have a clearer picture of the super-exchange versus double-exchange competition, we have performed the orbital decomposition of the exchange parameters as was done in Refs.~\onlinecite{korotin-jijs-Wannier,Fe-PRL}.

We have also performed an additional set of calculations in a supercell containing an actual spin-polaron and obtained qualitatively similar trends for exchange parameters as in case of homogeneous distribution of additional charge density. Although, the accuracy of those calculations was limited by a low number of $k$-points. The latter parameter is crucial for the calculation of the $J_{ij}$'s and therefore the results are not reported here. The most important outcome of these simulations is that the effective exchange interactions within the FM region were found to be substantially anisotropic as compared with VCA-derived results. 
The main reason of anisotropy around  $Mn^{3+}$ ions is attributed to Jahn-Teller distortion of the local ionic octahedral environment, which is not taken into account in a VCA calculation. 
The distortion results in the different population of the $x^2-y^2$ and $3z^2$ orbitals and hence the anisotropy of the magnetic couplings. 
This is also consistent with the anisotropy of the transition barriers, obtained in the series of calculation described in Sec III of the main text.
Apart from the more pronounced anisotropy, the obtained $J_ij$'s were qualitatively similar to the ones obtained with a simplified VCA calculations.

The thermal effects of the magnetic moments in the AFM system were studied solving The Landau-Lifshitz-Gilbert equation, as implemented in the UppASD package~\cite{skubic}, where each atomic magnetic moment, $\mathbf{m}_i$, is considered to be a three dimensional (3D) vector with constant magnitude

\begin{equation}
 \frac{\partial \mathbf{m_i}}{\partial t}=-\frac{\gamma}{1+\alpha^2} \mathbf{m_i} \times \mathbf{B}^\text{eff}_i-\frac{\gamma}{1+\alpha^2} \frac{\alpha}{m}\left[\mathbf{m_i} \times \left[ \mathbf{m_i} \times \mathbf{B}^\text{eff}_i\right]\right]
 \label{eq:LL}
\end{equation}
where $\gamma$ is the gyromagnetic ratio, $\alpha$ is the Gilbert damping parameter and $\mathbf{B}^\text{eff}_i$ is the effective field at the $i$-th site, which is defined as 

\begin{equation}
\mathbf{B}^{eff}_i=-\frac{\partial \mathcal{H}_\text{Heis}}{\partial \mathbf{m}_i}+ \mathbf{B}^\text{therm}_i(T).
\end{equation}
In this expression $\mathcal{H}_\text{Heis}$ is the Heisenberg Hamiltonian, containing all the interactions which model the system, and the second term is an stochastic field which introduces temperature effects.

The thermal field is modeled by using Gaussian white noise, which fulfills the following properties:

\begin{equation}
\langle  B^\text{therm}_i(T) = 0  \rangle \nonumber
\end{equation}
\begin{equation}
\langle B^{\text{therm},k}_i(t) B^{\text{therm},l}_j(t') \rangle=2D\delta_{ij}\delta_{kl}\delta(t-t'),
\end{equation}
where $D$ is the amplitude of the field $D=\frac{\alpha}{(1+\alpha^2)}\frac{k_B T}{\mu_Bm}$, where $i$ and $j$ are the lattice sites, $k$ and $l$ are the vector coordinates and $T$ is the temperature, respectively.

The calculations to estimate SP hopping barriers were carried out in the DFT framework using the VASP code \cite{Kresse1,Kresse2}. 
 We used the DFT+U methodology employing the Lichtenstein's approach\cite{lichtenstein} with $U$= 3.9 eV and $J$=0.9 eV for Mn $3d$ states. More details regarding our choice of Hubbard $U$ for this system can be found in Ref.\cite{Bondarenko3}.
The calculations of barriers were done for three cases: 1) bulk supercell; 2) 9 layer surface slab and 3) a diatomic layer consisting of 17 Ca, 18 Mn, 54 O, 1 La atoms ($x$=0.055). For 1) and 2) the unit cells consisted of 71 Ca, 72 Mn, 216 O and 1 La atoms, what corresponds to the La atomic fraction of $x$=0.013. For the layer and slab calculation the vacuum spacing was $\sim$18 \AA, large enough to isolate atoms from their periodic images.We used the PBE functional \cite{Perdew}, 550 eV as a cutoff energy and 2$\times$2$\times$2 k point mesh for bulk calculations and 2$\times$2$\times$1 k for slab and layer, respectively.
The obtained equilibrium  Pnma lattice parameters (a=5.29 \AA, b=7.44 \AA, c= 5.26 \AA) are in good agreement with the experimental data (a=5.28 \AA, b=7.46 \AA, c=5.27 \AA) \cite{lattice}.

The polaron motion is simulated by using an hybrid ASD-KMC algorithm. An usual simulations works by performing a standard ASD simulation, that is, each time step the LLG equation is solved, for a given magnetic configuration, with the \textit{polaron center}, i.e. the site that will define the ferromagnetic exchange cloud, located at a given site. The motion of the \textit{polaron center}, is determined by a KMC algorithm, which will calculate how long time, $\Delta t$, it will take for the \textit{polaron center} to move to another site. After a time $\Delta t$ has passed, the \textit{polaron center} is instantaneously moved to the new site, with the exchange interactions changing from AFM background to FM polaron region. 

Hence, of this way one can move the \textit{polaron center} making use of the energy barriers calculated from \textit{ab-initio}, while the magnetic texture evolves via the LLG equation, whilst tracking the time evolution of the exchange interactions, as given by the motion of the \textit{polaron center}.

The simulation of \textit{polaron center} motion was performed according to the following KMC algorithm, for time $t_\text{ASD}=0$ 

\begin{itemize}

\item[1]   Calculate the manifold $\{ r_{ji} \}$ of $N_i$ all possible transition  rates  from  state j to the initial state i. We assume  that the transition rate to NN or NNN spin-up (spin-down) (see fig. 3a) site finds according to the  Arrhenius law:

\begin{equation}
r_{ij}=\nu_0 e^\frac{-E_a}{K_B T}
\end{equation}

and $r_{ij}=0$ to all other sites (T  is the temperature and $E_{a}$ is the activation energy of the hopping event). 

The cumulative function then finds as $R_{ij}=\sum_j r_{ij}$,  where $j=1...N_i$.

\item[2] Take a random number taken as  $\rho \in  (0,R_{ij}]$). Select process j for which cumulative function satisfies the relation: $\sum_{j-1} r_{ik}< \rho \sum_{N_i} r_{ik}< \sum_{j} r_{ik}$

\item[3] Take a random number taken as  $\rho' \in  (0,1]$). Calculate the time needed for the system to evolve to the new state $\Delta t=-\frac{\log\left(\rho'\right)}{R_{ij}}$

\item[4]  Update the LLG equation from $t_\text{ASD}=t'$ until $t_\text{ASD}=t'+\Delta t$ then select state j. If $t_\text{ASD}=t_{max}$  end simulation,  else set $t'=t_\text{ASD}$ and go to step 2.

\end{itemize}



\section {II. The parametrized spin-polaron Hamiltonian}

\begin{figure}[htb]
\begin{center}
\includegraphics[scale=0.32]{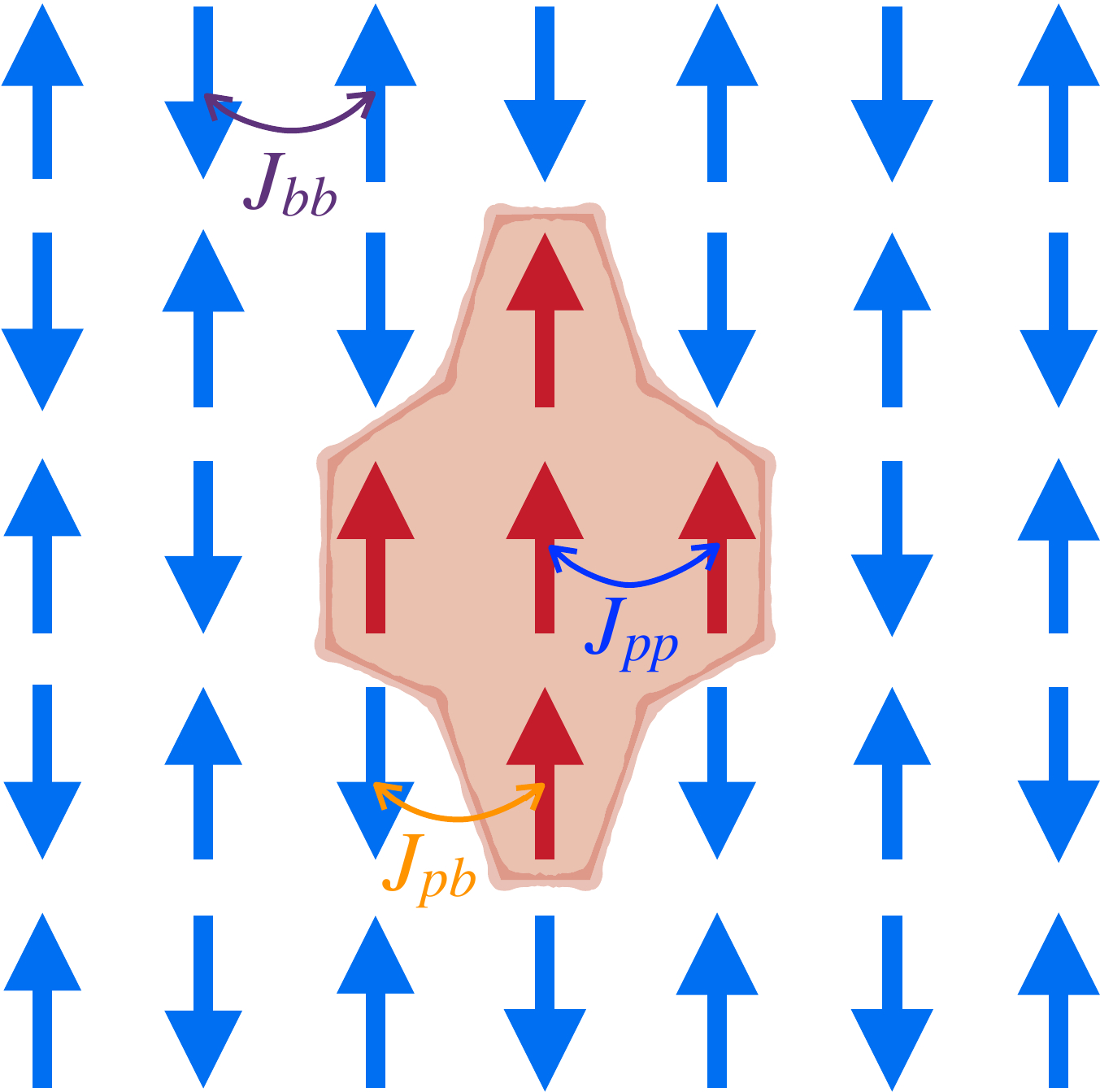}
\end{center}
\caption{(Color online) Schematic picture of the spin-configuration of ferromagnetic Mn atoms in the AFM background, defining the spin-polaron,  with three types of exchange interaction parameters. Mn moments corresponding to the spin-polaron are given in red colour.}
\label{fig-ASD} 
\end{figure}

For materials with a large magnetic moments, it is usually possible to map the magnetic configuration via the Heisenberg Hamiltonian
\begin{equation}
\mathcal{H}=-{1 \over 2}\sum_{i,j}J_{ij}\hat{s}_i\cdot\hat{s}_j
\label{heisenbergham}
\end{equation}
where the $J_{i,j}$'s are the Heisenberg exchange coupling parameters, which determine the strength of the magnetic interaction between the i-th and j-th magnetic moments $\hat{s}_i$. For spin-polarons there is a twist to this description, since electrons become localized over a few atomic sites, which in turn modifies local inter-atomic exchange interactions.
The model of the exchange interactions used here, is schematically shown in Fig.~\ref{fig-ASD}, where$ J_{bb}$ denotes the coupling between the spin corresponding to the AFM background of the CaMnO$_3$ lattice, $J_{pp}$ is the coupling within the spin-polaron body and $J_{pb}$ is the coupling of the boundary of the spin-polaron and the AFM background. Starting from the G-AFM reference state, which is the ground state for undoped CaMnO$_3$, we have computed the exchange integrals as a function of La doping. Details of these calculations were presented above. 

A critical evaluation of the accuracy of the obtained interactions are presented in the Appendix. Here a symmetry resolved analysis of the interactions show that the $t_{2g}-t_{2g}$ interactions are short ranged antiferromagnetic and rather insensitive to La doping. In contrast, the $e_{g}-e_{g}$ contribution is shown to be ferromagnetic in nature and strongly dependent on La concentration. For undoped $CaMnO_3$ the calculated exchange interactions, when combined with Monte Carlo simulations, reproduce with good accuracy the observed ordering temperature (see the Appendix). This establishes a level of accuracy of the simulations, and gives credence to using Eqn.\ref{heisenbergham} for the investigations of spin-polarons that are observed in La doped  $CaMnO_3$. As the Appendix shows, the thermal stability of these polarons is significant, of the same order as that of the underlying AFM order of $CaMnO_3$.

\section{III. Magnetic polaron hopping barriers}

As Fig.~\ref{fig-ASD} illustrates the spin-polaron is a region of ferromagnetic coupling in an AFM matrix, which is connected to the excess of electrons in this region. To study the dynamics of such polarons we adopt a concept where the spin-polaron can jump between two sites by overcoming a certain energy barrier, $E_a$. This idea refers to the well-known Marcus-Emin-Holstein-Austin-Mott (MEHAM) theory for polaron transfer \cite{Deskins}. Previously the method, albeit without magnetism, has successfully been applied to study lattice-polaron mobility at ionic interfaces and in band-gap semiconductors with low crystal symmetry \cite{Bondarenko1,Bondarenko2}

Studies based on ab-initio theory of the static properties of spin polarons in $La_xCa_{1-x}MnO_3$ have provided a theoretical description of the magnetic phase diagram in the La range of $0<x<0.10$, in good agreement with experimental data \cite{Bondarenko3}. These studies have shown that spin-polarons are stabilised mostly due to the magnetic interaction at lower La concentrations and due to the lattice contribution at larger concentrations. To reduce the influence of the spin-lattice correlations we chose here to calculate barriers for polaron hopping in the low La concentration limit, namely, for $x_{La}$=0.013. The barriers were estimated for hopping from the initial site to both nearest neighbor (NN) and next-nearest neighbor (NNN) sites, iin a process that is described in detail below.
We calculated the energy barriers for polaron hopping both for 3D and 2D geometry. 

In practice the polaron is formed from the antiferromagnetic matrix by flipping the magnetisation direction of one Mn atom, and then allowing the atomic positions and electronic structure to fully relax  (for an illustration see Fig 3a). This creates a local region of ferromagnetically coupled Mn atoms that contain one extra electron that becomes localized on the ferromagnetically coupled Mn atoms, thus forming a spin-polaron.
Spin-polaron hopping from an initial position to an adjacent location 
was controlled via the spin
configuration. Spins at two neighboring sites were simultaneously rotated by an angle $\gamma$ in a clockwise and anti-clockwise directions, respectively, and for each such configuration the electronic structure, magnetic moment and total energy was calculated using first principles theory (as described in the section with details of the calculations). This coordinated rotation of two Mn moments allows the ferromagnetic cloud to move through the lattice, and the extra charge associated to the spin-polaron was found to follow this cloud. Hence a simultaneous rotation of two Mn atoms was found to provide an excellent way to study the energy landscape of spin-polaron motion. For illustration we show in Fig.3a how a ferromagnetic region (indicated by thick bonds) is moved along the $<101>$ direction of the lattice. We refer to this movement as a nearest neighbor (NN) hopping, since there are no other Mn atoms located between the two Mn atoms that are rotated. We also investigated movement in the $<100>$ direction, and in this case the two Mn atoms that have their moments rotated have a third Mn atom between, with a fixed moment. For this reason we refer to motion along the $<100>$ direction as next nearest neighbor (NNN) hopping.
The energy barrier (or activation energy, $E_a$ ) of the polaron hopping is determined by the maximum of the total energy curve along the transition path (Fig.3b). Our results show that in each point of the transition path lattice relaxation lowers the energy with a non-negligible value as compared to the energy of the unrelaxed lattice (Fig 3b).

\begin{figure}[htb]

\begin{center}

\includegraphics[scale=0.33]{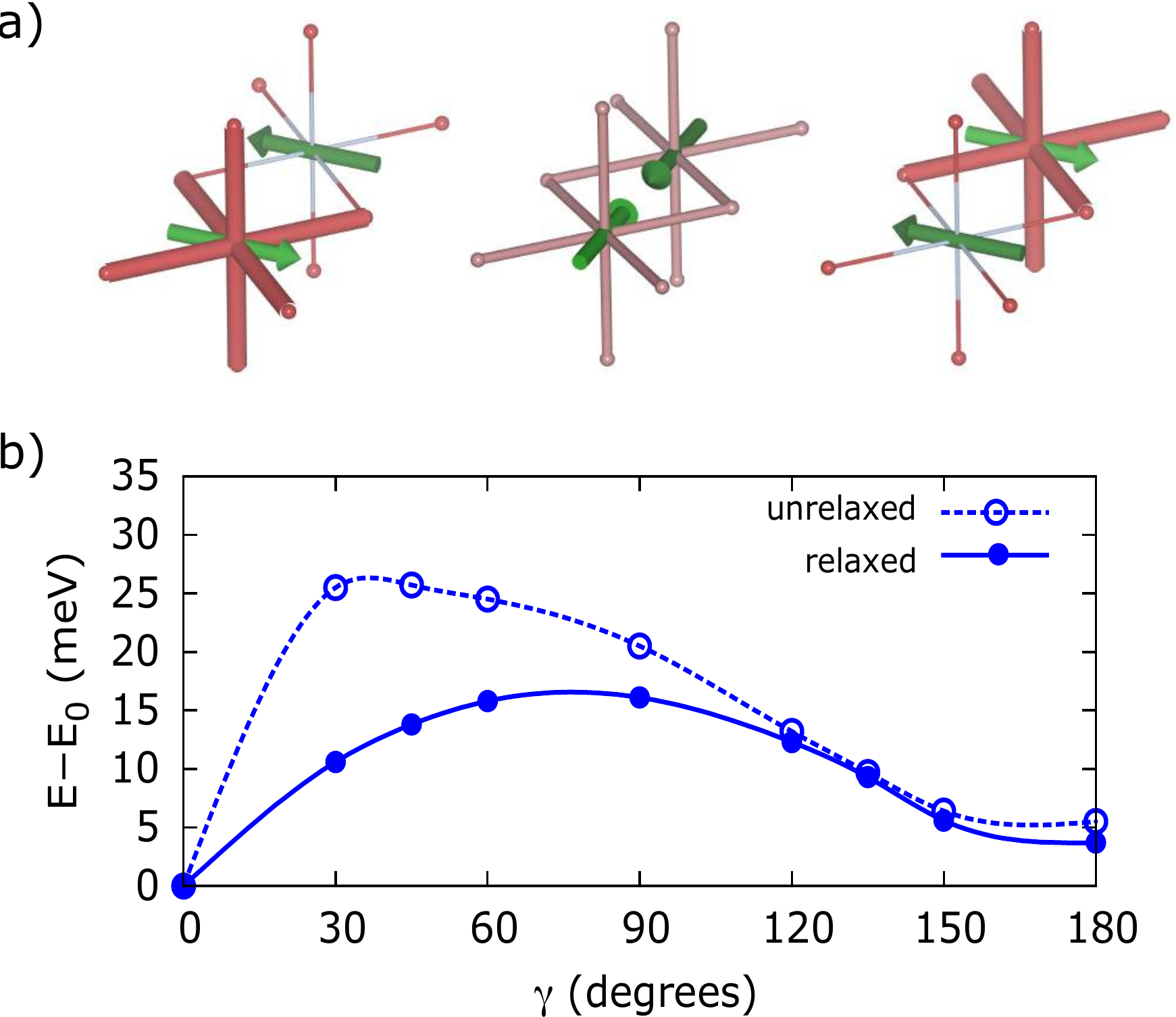}

\end{center}

\caption{\label{structure2}(Color online) Schematic illustration of the spin-polaron hopping mechanism.
a) Spin rotation angle of the initial Mn$^{3+}$ (solid red bonds) and final Mn$^{4+}$ (thin red bonds) sites, varied in the range of $\gamma=0^{\circ}-180^{\circ}$. Note that the moments are rotated clock-wise (anti-clock wise) for initial (final) sites. (b) Calculated transition barriers with suppressed (dashed line) and allowed (solid line) lattice relaxation. Energies of initial and final spin-polaron configuration differ due to the different distance between the spin-polarons and the La impurity. All energies along the transition path are given with respect to the ground state configuration (see also text).}

\label{fig5}

\end{figure}

The corresponding barriers were also calculated for free standing, 2D $La_xCa_{1-x}MnO_3$ layer and for the $La_xCa_{1-x}MnO_3$ surface (modeled as a slab). The maximum values of these energy barriers, together with the bulk data are collected in Table I. When a comparison can be made, the obtained energies agree well with experimental data and previous theoretical results. One can notice that the spin-polarons situated at the surface have slightly higher mobility (lower energy barrier) than those in bulk. It is also clear from Table I that the energy barriers for polaron hopping in different directions (labeled NN and NNN in Table I) are quite different. The lowest barrier was found for hopping to the NN site. Thus, we found that it was energetically favorable for the spin-polaron to move in the double-exchange active (001) plane. We also found that the transition barrier varied by a few meV depending on the La atom position with respect to the polaron. For this reason, in Table I we list two values for the bulk barrier.

\begin{table}[]
\centering
\caption{Spin-polaron hopping barriers obtained using DFT+U calculations. Data for a purely two-dimensional layer (Layer), the surface (Slab) and bulk (Bulk) of La doped $CaMnO_3$ are shown, both for nearest neighbour  (NN) and next-nearest neighbour (NNN). The lattice structure of the layer remained unstable during relaxation so that we show energies only for the unrelaxed case. Note that for bulk there are two sets of NN and NNN data, depending on the distance of the spin-polaron and the La doping atom. Earlier theoretical results from a t-J model are also listed. Experimental data explains hopping barriers obtained from the conductivity and resistivity measurements of $La_xCa_{1-x}MnO_3$ fitted to the adiabatic small polaron model.}
\label{Tb2}
\begin{tabular}{|l|c|c|l|}
\hline
\multicolumn{2}{|l|}{Unrelaxed lattice (meV)}                   & \multicolumn{2}{l|}{Relaxed lattice (meV)} \\ \hline
Layer, NN                         & 18                          & \multicolumn{2}{c|}{-}                     \\ \hline
Layer, NNN                        & 20                          & \multicolumn{2}{c|}{-}                     \\ \hline
Slab, NN                          & 21                          & \multicolumn{2}{c|}{10}                    \\ \hline
Slab, NNN                        & 23                          & \multicolumn{2}{c|}{14}                    \\ \hline
Bulk, NN                          & 24-27                       & \multicolumn{2}{c|}{14-18}                 \\ \hline
Bulk, NNN                       & 44                          & \multicolumn{2}{c|}{33}                    \\ \hline\hline
\multicolumn{2}{|l|}{t-J model, NNN \cite{Meskine2}} & \multicolumn{2}{c|}{40} \\ \hline
\multicolumn{2}{|l|}{Experiment, \cite{Ohtaki, Lan, Worledge} (x=0.02)} & \multicolumn{2}{c|}{20} \\ \hline
\end{tabular}
\end{table}

\section{IV. Magnetic polaron motion}

\begin{figure}[htb]
\begin{center}
\includegraphics[scale=0.23]{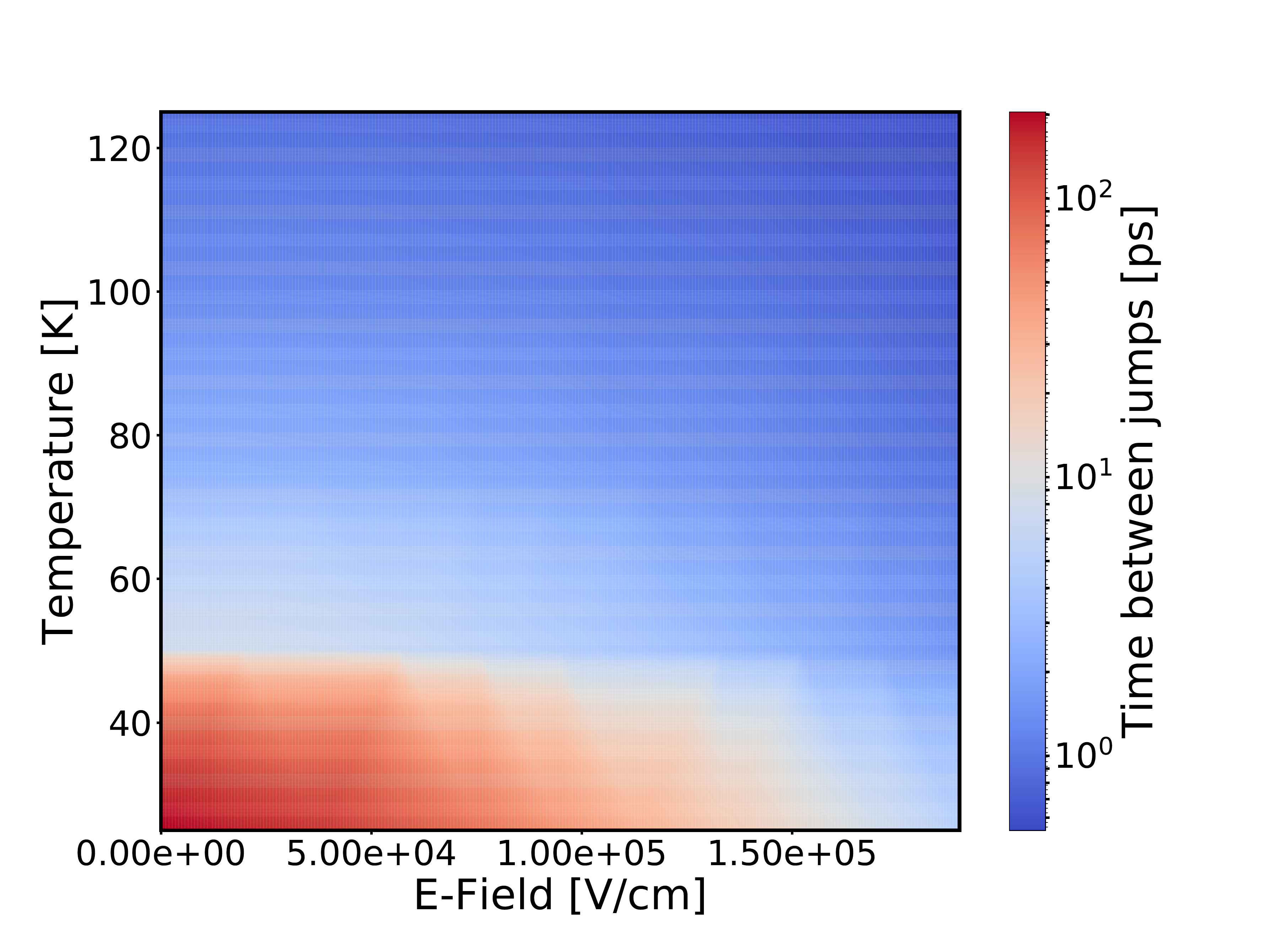}
\end{center}
\caption{\label{structure3}(Color online) Average times between spin-polaron jumps of La doped $CaMnO_3$, obtained from the simulations (see text) for different temperatures and strengths of an applied E-field.}
\label{fig6} 
\end{figure}

\begin{figure*}[htb]
\begin{center}
\includegraphics[scale=1.40]{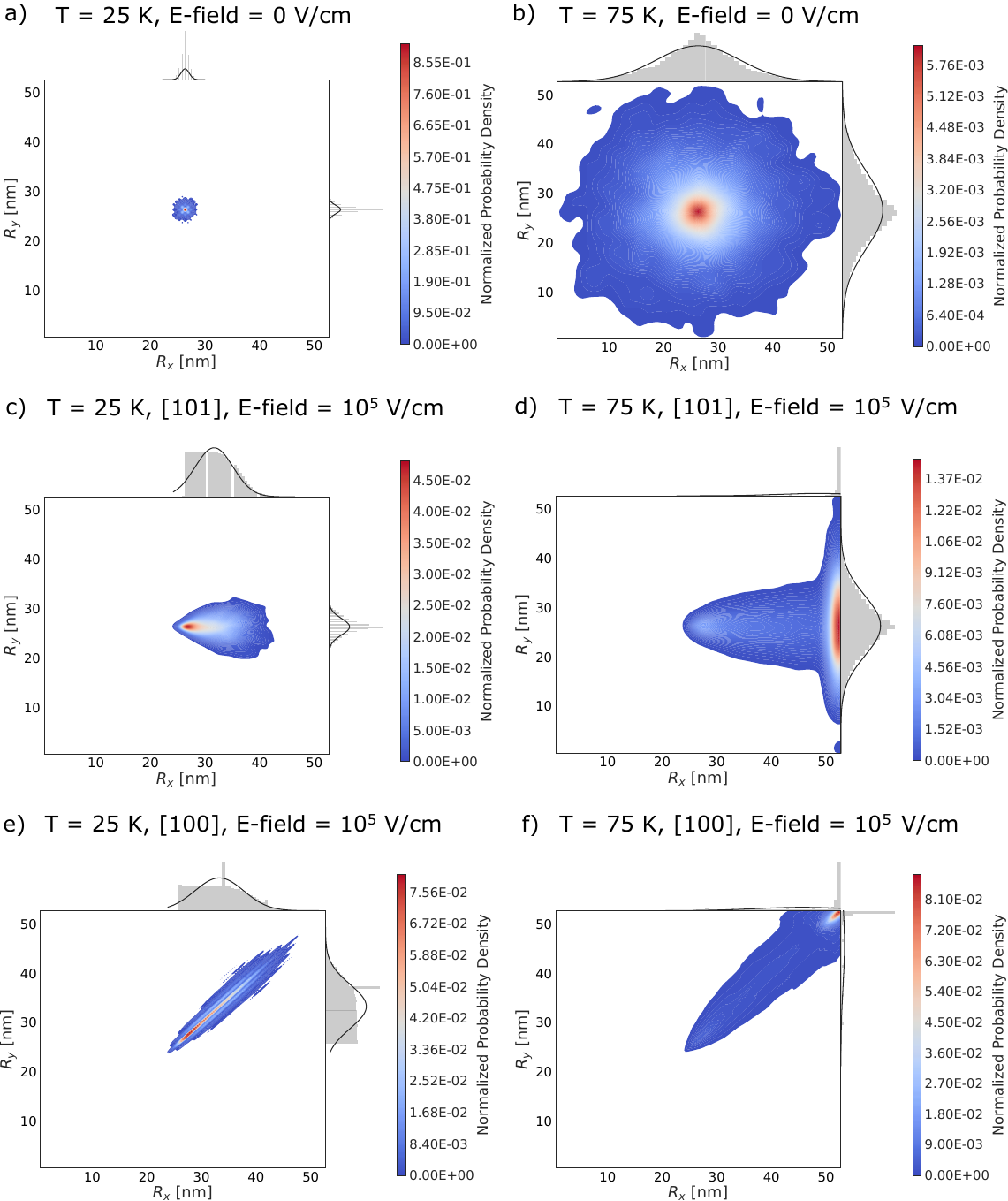}
\end{center}
\caption{\label{structure4}(Color online) Probability density of single spin-polaron propagation at variable temperatures and E-fields. The probability densities were obtained during a simulation time of 1 ns. The probability is given both as a color code (illustrated by the color bar shown on the right of each figure) and as a histogram projected on positions in x- and y-direction (shown to the right and the top of each figure).}
\label{fig7} 
\end{figure*} 

\begin{figure*}[htb]
\begin{center}
\includegraphics[width=0.75\textwidth]{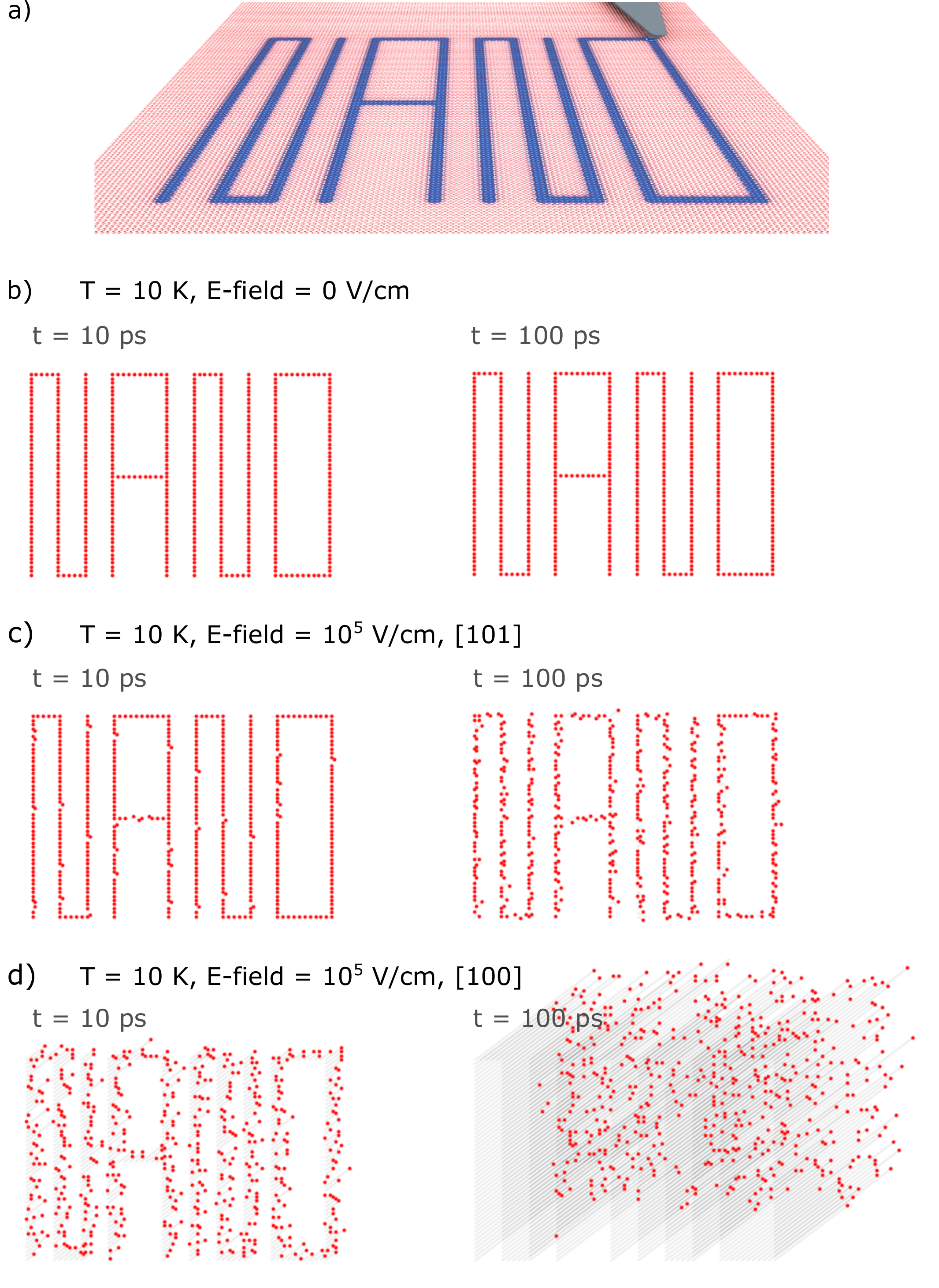}
\end{center}
\caption{\label{structure5}(Color online) a) Cartoon showing schematically how electrons can be introduced into a substrate of  $CaMnO_3$, by an STM tip, to store information via the formation of polarons that are immobile if no external stimuli, like an E-field, are present (picture kindly provided by Ms Ella Kurland). Simulation for multipolaron system with NN and NNN hoppings allowed in the system for b) E=$0~V/cm$, c) E-field=$10^5$~V/cm in $\left [100\right ]$ direction and  d) E-field=$10^5$~V/cm in $\left [101\right ]$ direction.}
\label{fig8} 
\end{figure*}

In this section, we discuss magnetic temperature induced motion of the spin-polaron. The hopping rate was obtained from an Arrhenius-type process with the activation barrier $E_a$.  The magnetic propagation from site to site was modeled using a hybrid ASD-KMC  method. In this algorithm, the polaron can be characterized by two quantities, the \textit{polaron center}, that is the site where the electron is self-localized and the actual \textit{polaron texture}, i.e. the actual ferromagnetic texture resulting from the local ferromagnetic interactions. At each time step the Landau-Lifshitz-Gilbert equation is used to determine the dynamics of the magnetic moments (AFM background and \textit{polaron texture}). Whilst, the KMC part of the algorithm calculates the time it takes for the \textit{polaron center} to move to a neighboring site, in which case the exchange interactions change instantaneously.
For each temperature,  we have performed 100 different simulations with different seeds for the random number generator to obtain statistically relevant results. The energy barrier of the hopping process to the NN and NNN sites has been set as the average values obtained by the first principle calculations for the 3D case, as reported above (see Table I). 
The attempt frequency, $\omega_0$, was set to be $\omega_0=1\times10^{12} \text{ Hz}$, a value that  is expected if we assume spin-polaron motion driven by magnonic and phononic processes \cite{Dolling,Kaplyanski}. 

We evaluated the spin-polaron hopping process using the hybrid ASD-KMC algorithm and tracked average times between \textit{polaron-center} jumps, both in the case of the random motion (no electric field applied) and in the presence of an applied external E-field. The effects of an applied E-field was introduced with a term $q\mathbf{E}\cdot\mathbf{r}$ ($q$ is electron charge and $\mathbf{r}$ the position of the \textit{polaron center})  to the energy barrier used in the simulations, taking into account both electronic and magnetic contributions to the energy. 

We observed from the simulations that both the temperature and strength of the E-filed influenced the mobility of the polaronic center, and the results of the time between jumps of the \textit{polaron center} from one site to the next is shown in Fig.4. It may be seen in this figure that the average times between polaron jumps decreases by about two orders of magnitude from $~$ 100 ps to $~$ 10 ps as the temperature increases from 25 K to 125 K.  From this figure it is also clear that the hopping rate of the spin-polarons changes when an external electric field is applied. For instance, at 25 K the jump time changes with about one order of magnitude when the jump rate reaches $\sim 2 \cdot 10^5$ V/cm.  

We have furthermore followed the position of a \textit{polaron center}, as it  moves through the lattice during the KMC simulation. The normalized probability density of detecting a \textit{polaron center} at a particular position of the lattice is plotted in Fig. 5, as a function of both temperature and applied E-field. Note that the probability density shown in Fig.5 represents all the events that occurred for the average of 100 simulations, each one covering dynamics over a total time of 1 ns.
In the absence of an E-field, the spin-polaron performs a random walk motion, that becomes more diffuse, the larger the temperature is (Fig.5a,b).
Using the obtained maximum average distances at different temperatures,  we have estimated the polaron diffusivity coefficient according to random-walk diffusion model as $D=\frac{\left \langle R^2 \right \rangle}{4t}$, where $\left \langle R^2 \right \rangle$ is the average maximum distance which the \textit{polaron center} has passed, and $t$ is the simulation time \cite{Bressloff}. The obtained values of $D$ lie in the range of  $0.5\cdot10^{-8}-1.5\cdot 10^{-6} m^2/s$ (25-125 K), and can be compared to characteristic values for polaron diffusion processes \cite{Coehoorn}. When an external E-field is applied, the polaron moves along this field (Fig.5c-f). For a given strength of the E-field, an increased temperature makes the polaron movement faster, which is natural since the spin-polaron movement is an activated process. It is notable, that lowering of the activation energy by the external electric fields has been recently proved experimentally in manganese oxide heterostructures \cite{Kuang}.  

Note that in the chosen parameter range of temperatures and E-field, it is quite possible to control the position and the speed of the polaron. For instance, in the absence of an applied E-field and at a temperature of 25 K or 75 K, the polaron does not move significant distances, and stays instead at its original position (Figs.5a and 5b). However, an applied E-field of $10^5$ V/cm can easily move the polaron over significant distances. For instance, an E-field of $10^5$ V/cm in the [100] direction moves (over a time of 1 ns) the spin-polaron a distance of some 10 nm, when T=25 K (Fig.5c) and over 30 nm when T=75 K (Fig.5d). An E-field in the [101] direction is seen to induce even larger distances of the spin-polaron motion (Figs.5e and 5f). This means that he drift velocity of spin-polarons is of order 10-30 m/s. Note that this speed should be distinguished from the speed of actual individual spin-polaron motion, or the spin-polaron mobility, that is two to three orders of magnitude larger. 

 \section{V. Spin-polaronics as a technology}

The section above shows that the thermal stability of the spin-polaron in La-doped  $CaMnO_3$ provides an excellent opportunity to use these quasiparticles in technological applications. As the results above show, a polaron can both be kept in a fixed position for a given period of time, and made to move in desired directions by a rather weak applied E-field. Hence we propose that storing and erasing information by means of spin-polarons in $CaMnO_3$ should be rather straight forward, something this section outlines. The basic idea is that electrons added to a substrate of $CaMnO_3$ form polarons, that stay put for any period of time, and only after an applied electric field are they moved in any desired direction. By dressing the electron with a magnetic cloud, this procedure allows to make the motion of a charged quasiparticle more 'classical' compared to undressed free charges. 

The extra electrons can either be introduced via doping with a trivalent atom (e.g. La) on the Ca site of $CaMnO_3$, or as Fig. 6 schematically shows, via an external source, e.g. as provided by a STM tip. Since these extra charges are associated with a magnetic contrast (ferromagnetic) that is different compared to that of the background material (antiferromagnetic), their detection is straight forward from AFM or STM techniques \cite{Wiesendanger}. Hence it is possible to store information, e.g. as Fig.6 shows, in the form of text written with letters in nano-size, simply by adding electrons in an appropriate pattern. This information can be stored or erased at any time, simply by application of an external E-field.

To illustrate this possibility outlined above, we have performed KMC evolution of the multipolaron system schematically shown in Fig. 6. We started by introducing into a  $CaMnO_3$ substrate extra electrons in a controlled way, so that formation of polarons, outlined with a magnetic contrast the text ''\texttt{NANO}''. 
If the temperature is sufficiently low, this text stays intact for a period larger than 100 ps at 10K (see Fig.6b). However, as Figs.6c and 6d show, an applied E-field erases this text without difficulty. Movies illustrate in real time how the text ''\texttt{NANO}'' is robust against thermal fluctuations, but can be erased when an external E-field is applied\cite{Supplementary}. This form of writing and erasing text is a natural, but extreme miniaturization of storing information, and it is conceivable that further condensation of information is technologically impossible. It should be noted that writing and erasing text in nano-sized letters has by now been realized many times, starting with the pioneering results of Ref.\cite{Eigler} that used Xe atoms to spell out the name ''\texttt{IBM}''. However, the technology proposed here does not suffer from extremely long writing and detection times, in contrast to Ref.\cite{Eigler}.

\section {Conclusion}

In this work we have investigated theoretically the static and dynamic properties of spin-polarons in La-doped $CaMnO_3$. In order to do this, we constructed an effective low energy Hamiltonian, in which all parameters were calculated from first principles theory. This low energy Hamiltonian is used to investigate the temperature stability of the spin-polaron, as well as the response to an external applied electric field. Technically this involves ab-initio electronic structure theory and atomistic spin-dynamics simulations in combination with kinetic Monte Carlo simulations. In our study we compared results from different geometries, like spin-polarons in bulk, surface and as single two-dimensional layers, and significant differences were observed. Where a comparison can be made, primarily for bulk geometries, the results presented here compare well with experimental data, and previous theory. 

We demonstrated a remarkable control of the mobility of spin-polarons in this material, and that the critical parameters deciding this, is the temperature and the strength of the applied electrical field. This opens up for technology using spin-polarons, and our simulations demonstrate that storing and erasing information magnetically, by introduction and control of electrical charge, is possible, even for rather low strength of the external E-field. We demonstrate that it is possible to write text in atomic sized letters, similar to the pioneering text written by atoms that were moved around with an STM tip\cite{Eigler}. The advantage with the technology proposed here, is the vast speed with which information can be stored and erased. It is tempting to contrast the information density, and writing speed, proposed in Fig.6, to that of the over 2000 thousand year old technology behind the Rosetta stone.

The technological implications of electrical control of spin-polarons is only touched upon in the present work. It is foreseeable that other technologies may be just as relevant. We propose that transistor functionality of spin-polarons, e.g. in $CaMnO_3$, is possible, given their stability and the ease with which one may control their movement with an electric field. Hence charges injected at one end of a device built from $CaMnO_3$ can be moved from a source to a drain, and be controlled by an E-field provided by a gate. Such studies are underway. Also, exploration of a wider selection of host materials that can host spin-polarons is interesting, both to establish functionality at even higher temperature, but also to find materials where a significant Dzyaloshinskii-Moriya interaction plays role, so that potentially spin-polarons with unique chirality can be stabilized. It is foreseeable that in light of the results shown here, spin-polaronics may enter an era with many new and exiting results in basic science and technology.

\section {Acknowledgments:}

\begin{acknowledgments}
We acknowledge the financial support by the eSSENCE, the Swedish Research Council and the KAW foundation (projects 2012.0031 and 2013.0020). The computer simulations were performed on resources provided by the Swedish National Infrastructure for Computing (SNIC) at the National Supercomputer Centre (NSC) and High Performance Computing Center North (HPC2N).
\end{acknowledgments}

\section {Appendix: Static properties and thermal stability of the spin-polaron}

Computing the exchange parameters for the system containing a localised spin-polaron is a computationally demanding procedure, due to the large size of the simulation cell. 
Here, in order to model the La doping of  $Ca_{1-x}MnO_3$, we have used the Virtual Crystal Approximation (VCA) to take La doping into account. This approach is natural in introducing free charge carriers in the valence band, and has sufficient computational efficiency. In addition, it was shown for doped LaMnO$_3$\cite{solovyev-LMO-VCA}, that VCA is able to reproduce the physics of the double-exchange mechanism in a relatively wide concentration range. We expect, that it is particularly applicable in the low-doping regime, since the modification of the electronic structure is expected to be small.  

\begin{figure}[htb]
\begin{center}
\includegraphics[scale=1.00]{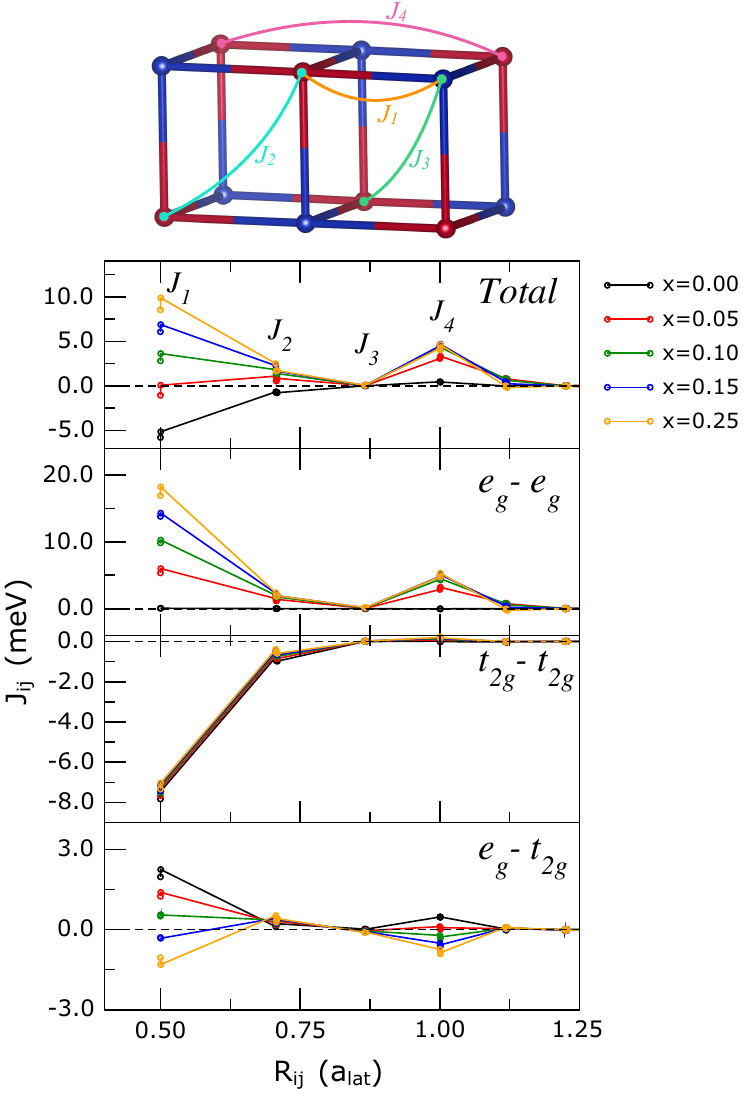}
\end{center}
\caption{(Color online)  Computed exchange parameters between Mn atoms in La doped CaMnO$_3$ for different La concentrations, $x$, and as function of inter-atomic distance (in units of lattice constant a). The definition of $J_1$,  $J_2$,  $J_3$ and  $J_4$ is illustrated in the top of the figure.}
\label{jijs} 
\end{figure}

The obtained results of exchange interactions (Fig.~\ref{jijs}) show that the $t_{2g}-t_{2g}$ contribution is rather short-ranged and of AFM nature, which is typical for the super-exchange mechanism with Mn-O-Mn bonds that have angles close to 180 degrees. Moreover, this channel of the exchange interaction is independent of La concentration. As Fig.~\ref{jijs}  shows, there is a small $e_g-t_{2g}$ term, which appears due to the mutual tilting of MnO$_6$ octahedra (for undistorted octahedra this channel would be zero, by symmetry \cite{Cardias}). The $e_g-e_g$ contribution to the exchange integrals is zero for $x=0$ due to that these orbitals are empty in this situation. However, upon doping these states become populated and start to participate in the magnetic interactions, predominantly with a FM character, due to their double-exchange nature. As Fig.~\ref{jijs}  shows, the  $J^{e_g-e_g}$ exchange is relatively long-ranged and a close scrutiny of these results show that the nearest neighbor interaction is almost linearly proportional to the La doping.

Fig.~\ref{jijs} shows that $J_1$ and $J_4$ (defined in Fig.~\ref{jijs}) are the exchange parameters that most strongly affected by La doping.
This agrees with previously reported results for La-doped CaMnO$_3$ ~\cite{solovyev-LMO-VCA}.
These are the interactions between the atoms belonging to the same Mn-O-Mn-chains. Along these directions, the lobes of the $e_g$ orbitals are oriented towards O-p states, that facilitates the electron hopping process and the double exchange mechanism.

The calculated $J_{ij}$'s together with the magnetic moments (that we calculated to be 2.51 $\pm$ 0.05 $\mu_B$/Mn atom, for all La concentrations), allow us to parameterize the inter-atomic exchange interactions of the spin-polaron, via an effective spin-Hamiltonian. When combined with atomistic spin dynamic (ASD) simulations, this opens up for an investigation of the dynamics of these objects. 
In order to model the magnetic interactions of the spin-polaron, one has to take into account that charge gets localized in space, and we made the following treatment to simulate this: 

(i) First of all, we assume that the $e_g$ electron is localized within the body of the polaron and hence does not participate in interactions with long range that goes outside the spin-polaron. This implies that we considered only NN exchange couplings for the ASD simulation.
This approximation was tested against the ab-initio calculated distribution of the excess charge associated with the seven-site spin-polaron in bulk. We found that the electron associated with the spin-polaron is distributed essentially only over a central atom and its nearest neighbours, which justifies a short ranged double exchange within the spin-polaron only.

(ii) Second, we approximate the distribution of the excess $e_g$ electron to be homogeneous over the central site and its neighbours in the magnetic polaron. 
In this way, the interactions in the inner part of a spin-polaron, with $N_s$-sites, 
is described by the set of $J_{ij}$'s from Fig.~\ref{jijs} corresponding to $x=1/N_s$. On the boundary of the spin-polaron, the exchange coupling occurs between the ions with $t_{2g}^3$ and $t_{2g}^3e_g^{1/N_s}$ configurations. 
In this case, the $e_g-t_{2g}$ contribution can be approximated as $J^{e_g-t_{2g}}_1(x=1/(2N_s))$, due to the linear dependence of this coupling on $x$. 
Summarizing, we arrived at an interpolation scheme of exchange interactions of spin-polarons of any size, defined from the calculated nearest neighbor parameters shown in Fig.~\ref{jijs}, in the following way:
\begin{gather}
J_{pp} = J_1(x) \\ 
J_{pb} = J^{t_{2g}-t_{2g}}_1 (x/2) + J^{e_g-t_{2g}}_1(x/2)  \\
J_{bb} = J_1(x=0).
\end{gather}
We have used the relationship $x={1 \over {N_s}}$, since one electron is associated with any $N_s$-site polaron.
According to the formalism introduced above, the parameters listed in Table~\ref{Tb1} have been used for the dynamics of the spin-polaron. 

\begin{table}[thb]
\caption{Parameters used for the ASD simulations for $N_s$-site polarons. $M_p$ refers to the values of the Mn magnetic moments belonging to the polaron. $M_b$ corresponding to the background spins was calculated to be2.45 $\mu_B$. The coupling $J_{bb}$ was calculated to be 5.35 meV. The exchange parameters $J_{pb}$ and $J_{pp}$ are defined in the text.} 
\centering
\begin{tabular}{cccc}
\hline
$N_s$ & $M_p$ ($\mu_B$) & $J_{pp}$ (meV) & $J_{pb}$ (meV) \\
\hline
5   & 2.56 & 9.40 &  -6.77 \\ 
7   & 2.53 & 6.61 &  -6.54 \\ 
8   & 2.52 & 4.99 & -6.31 \\ 
11 & 2.51 & 3.36 & -6.09 \\ 
13 & 2.50 & 1.99 & -5.85 \\ 
\hline
\hline
\end{tabular}
\label{Tb1}
\end{table}

\begin{figure}[htb]
\begin{center}
\includegraphics[scale=0.55] {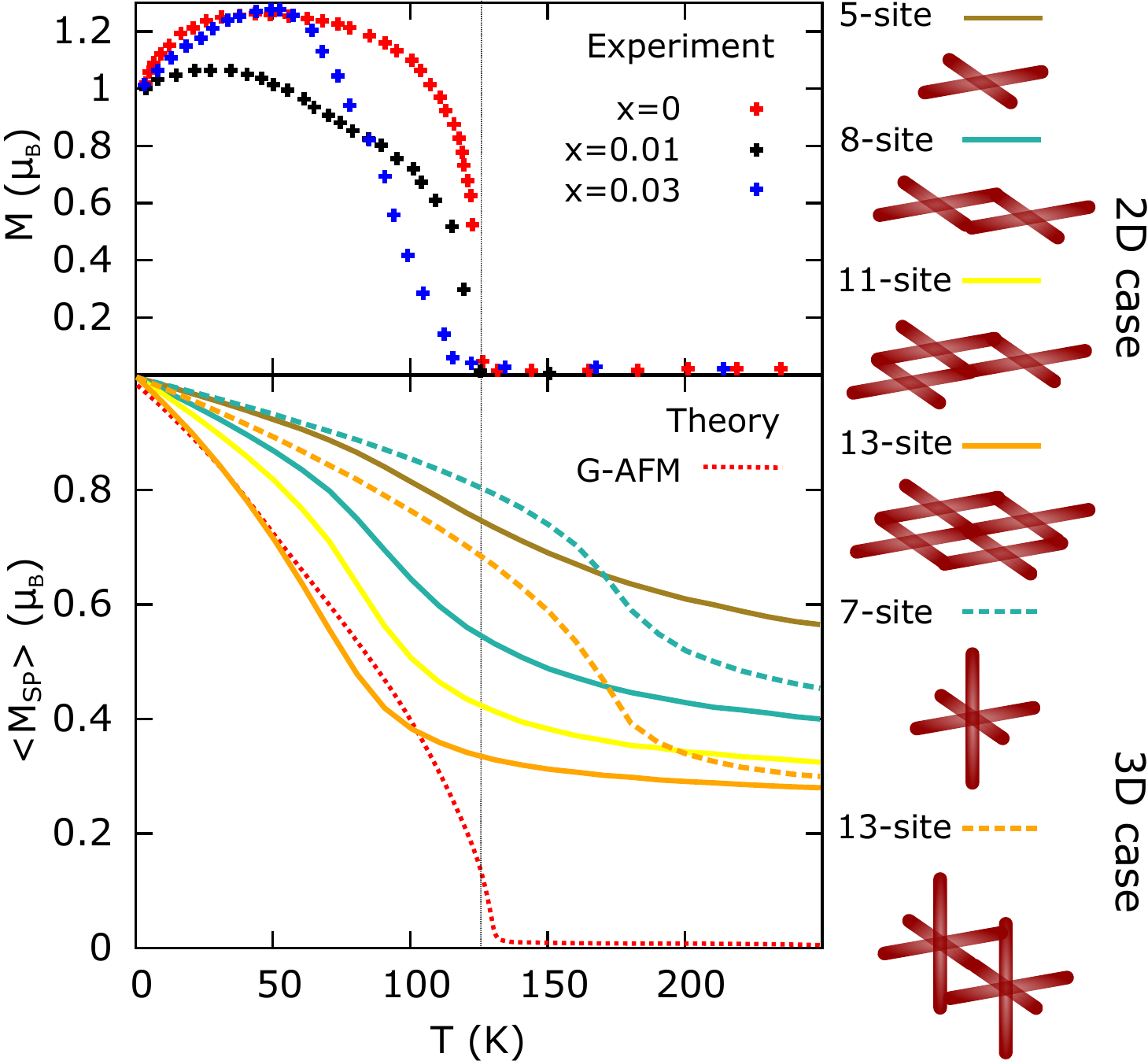}
\end{center}
\caption{\label{structure6}(Color online) Top panel. Crossed points indicate experimentally measured magnetisations in the bulk sample,  for lanthanum doping levels of x=0.01-0.03 \cite{Cortes}. Note that experimental magnetisation curves are normalised to 1 at 0 K. Bottom panel. Computed average magnetisation for a variety of spin-polaron configurations in 2D and 3D lattices (schematically shown to the right), using an effective spin-Hamiltonian and Monte Carlo simulations (see text). Results for the 3D case are shown with dashed lines while the results of the 2D case are shown as solid lines. Note that different sizes of polarons are investigated. The experimental ordering temperature of bulk $CaMnO_3$ ($\sim$125 K) is indicated by a vertical dashed line.}
\label{fig4} 
\end{figure}

Next we used the extracted exchange parameters to study the stability of the magnetic polarons, with respect to temperature fluctuations. The investigation is done both for a 2D and 3D lattice, making use of an extended Heisenberg Hamiltonian
\begin{eqnarray}
\label{HH}
\mathcal H = -{1 \over 2}\sum_{i,j} J_{ij} \hat{m}_i \cdot \hat{m}_j-K_\text{ani}\sum_{i}\left( \mathbf{e}_i\cdot \mathbf{e}_\textrm{K}\right)^2 ,
\end{eqnarray}
with the exchange parameters listed in Table 
\ref{Tb1}. Here \textbf{m$_i$} denotes the Mn-projected atomic magnetic moment at site $i$. Furthermore,  $K_{ani}$ is the parameter characterizing the magnetocrystalline anisotropy, while $\mathbf{e}_i$ and  $\mathbf{e}_\textrm{K}$ are the direction of the moment of atom $i$ and the easy axis direction, respectively. The magnitude of the $K_\text{ani}$ is set to $0.01\text{ mRy}$, and it is used to have a well defined quantization axis, mostly for visualization purpouse. In general, the presented results are not strongly dependent on the value of the anisotropy. In these calculations it is only the thermal stability of the magnetic sublattice that is of interest, something we investigated by means of Monte Carlo simulations, using the Metropolis-Hastings algorithm, of the spin-system, where the polarons were considered to be fixed at a given crystal site, and were not allowed to move from site to site. We considered magnetic polarons of varying size, in the range of 5-13 sites (7-13 sites) for the 2D (3D) case, and the polaron geometries outlined in the right-hand side of Fig.\ref{fig4}. To improve the statistics for each spin-polaron configuration, one hundred different realizations, with different random number seeds of the Monte Carlo simulations had been taken. 

The average magnetisation of the polaron region, $\langle \mathbf{M}_\text{SP}\rangle$ has been considered  as the average magnetization taken over the polaron region, $SP$, and averaged over the different realizations:

\begin{equation}
\langle \mathbf{M}_\text{SP}\rangle=\frac{1}{N_\text{ens}}\sum^{N_\text{ens}}\frac{1}{N_{max}}\sum_{n\in SP}^{N_{max}}\mathbf{m}^n
\label{eq:pol_stab}
\end{equation}

For the treatment of 2D systems, a 100x100 square lattice was evaluated as a basis for the simulations, where periodic boundary conditions were considered in the in-plane directions (for 3D systems we have considered supercells of  20x20x20, with periodic boundary conditions in all directions).  

In Fig.\ref{fig4} we show the average magnetisation for different spin-polaron configurations. It is interesting first to note that calculations performed for impurity-free $CaMnO_3$ indicate a critical temperature at $\sim$ 135 K in excellent agreement with the experimental data (Fig.\ref{fig4}, top panel). This establishes the accuracy of the exchange parameters and the spin-model used to investigate temperature effects. The values of the magnetisation of the spin-polaron region is seen to be non-zero well above the ordering temperature for all choices of geometry of the spin-polaron. This deonstrates the ropustnes of the spin-polaron with respect to thermal fluctuations. It is however not a reflection of exchange interactions being stronger in this region, but rather a consequence of short range ordering that survives in most magnets well above the bulk ordering temperature. 
It can also be seen in Fig.\ref{fig4} that for the 2D case, $\langle M_{SP} \rangle$ decreases quicker as a function of temperature  (Fig.\ref{fig4}, bottom panel) compared to the magnetisation curve of bulk  $CaMnO_3$, as expected for a system with reduced coordination number. We also notice that for the 2D case, the temperature dependence of the spin-polaron is almost independent on size. 

As for the 3D magnetisation (see Fig.\ref{fig4}, bottom panel, dashed lines) we observe similar trends to that in the 2D lattice. However, one may notice that the magnetization for all magnetic polaron configurations saturates at higher temperatures than in the 2D, as expected. We also note that for the 3D case, the size dependence is weak when it comes to the behaviour of $\langle M_{SP} \rangle$ with respect to temperature. Interestingly, for the 3D polarons,  $\langle M_{SP} \rangle$ remains finite well above the ordering temperature of undoped  $CaMnO_3$ ($\sim$135 K), as is shown in Fig.\ref{fig4}. 
The result agrees with experimental results reporting stability of the polaronic centres in the paramagnetic phase beyond the Neel temperature \cite{Chiorescu}.

The results presented above indicate the importance of the coordination number for spin-polaron stability. As we already mentioned in the introduction, previous theoretical studies on spin-polaron formation in lanthanum doped $CaMnO_3$ discovered charge localisation preferably in the double-exchange active plain (101) \cite{Bondarenko3}. However, spin-polarons with higher coordination number (in bulk against of surfaces, for example) seem to stabilise at higher temperatures.

\end{document}